# The second phase transition in the pyrochlore oxide $Cd_2Re_2O_7$

Zenji HIROI*, Jun-Ichi YAMAURA, Yuji MURAOKA, and Masafumi HANAWA

Institute for Solid State Physics, University of Tokyo, Kashiwanoha, Kashiwa, Chiba 277-8581, Japan

Evidence for another phase transition at 120 K in the metallic pyrochlore oxide $Cd_2Re_2O_7$, following the structural transition at 200 K and followed by the superconducting transition at 1.0 K, is given through resistivity, magnetoresistance, specific heat, and X-ray diffraction measurements. The results indicate unique successive structural and electronic transitions occurring in the pyrochlore compound, revealing an interesting interplay between the crystal and electronic structures on the itinerant electron system in the pyrochlore lattice.

*E-mail address: hiroi@issp.u-tokyo.ac.jp

$Cd_2Re_2O_7$ is the first and only one superconductor ($T_c$ = 1.0 K) in the family of pyrochlore oxides which at $T_{s1}$ = 200 K where the temperature dependence of resistivity and magnetic susceptibility changes markedly.[1, 4, 5] It accompanies a structural transition which was suggested to be from the ideal cubic pyrochlore structure with space group *Fd*3*m* to another cubic structure with space group *F*43*m*.[4] However, recent Re NQR experiments have indicated the lack of three-fold axis below $T_{s1}$, implying that the true symmetry is lower than cubic.[6] The structural determination of the low-temperature phase has not yet been completed.

Besides the transition at $T_{s1}$ it was noticed in our previous study that the resistivity curve showed a tiny anomaly around 120 K.[1] In this letter we present strong evidence for another phase transition at $T_{s2}$ ~ 120 K through the measurements of resistivity ($\rho$), magnetoresistance ($\Delta\rho$), specific heat ($C$), and X-ray diffraction (XRD). The results reveal interesting successive phase transitions in the compound: the highly symmetrical pyrochlore lattice is deformed slightly in two steps, and, at the same time, remarkable changes in the electronic structure occur correspondingly, and finally superconductivity sets in as a ground state for the slightly distorted pyrochlore structure.

Single crystals of $Cd_2Re_2O_7$ prepared as reported previously[1] were used for all the measurements. Resistivity was measured by the standard four-probe method in a Quantum Design PPMS system. Specific heat measurements were carried out by a heat-relaxation method also in the PPMS system. Magnetic susceptibility was measured in a Quantum Design MPMS system. Single-crystal XRD experiments were performed in an Imaging plate type diffractometer down to $T$ = 10 K and a four-circle diffractomer down to $T$ = 85 K.

As reported previously, the resistivity of $Cd_2Re_2O_7$ shows an unusual independence of temperature above

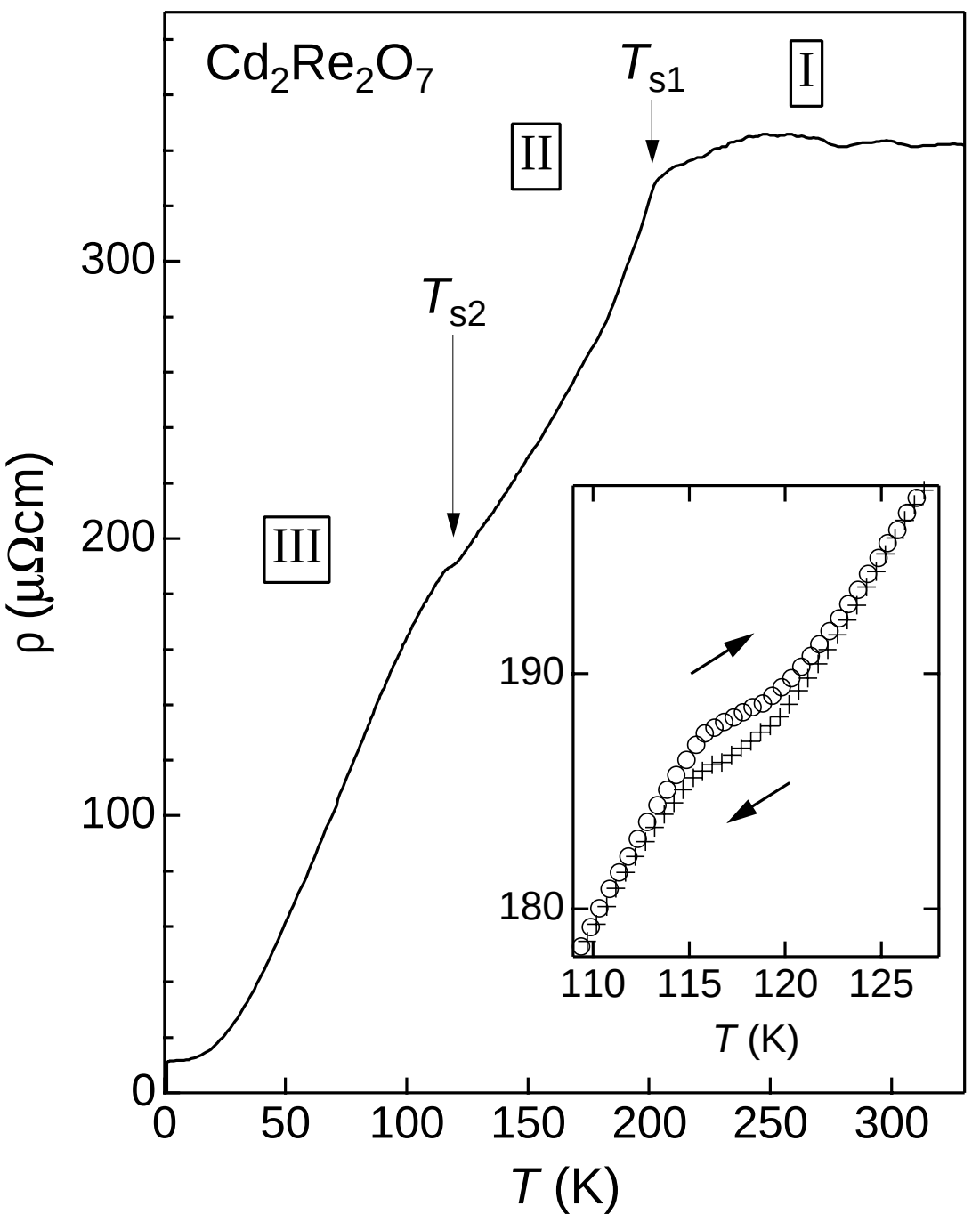

Fig. 1. Electrical resistivity ($\rho$) of a $Cd_2Re_2O_7$ single crystal showing two anomalies at $T_{s1}$ = 200 K and $T_{s2}$ = 120K. The measurements around $T_{s2}$ shown in the inset were carried out, after slow cooling, on heating (open circle) and then on cooling (cross) with a rate 0.1 K/min. A superconducting transition occurs at $T_c$ = 1.0 K.

$T_{s1}$ = 200 K, while it suddenly decreases below $T_{s1}$. Figure 1 shows a typical resistivity curve for a high-quality crystal with residual-resistivity-ratio more than 30. To be noted here is a small anomaly seen at $T_{s2}$ = 120 K, as shown in the inset to Fig. 1, which may indicate a signature for a phase transition. Both on cooling and heating there are two knees at $T$ ~ 116 K and 119 K with plateaus in-between. A thermal hysteresis is seen clearly between the two curves, suggesting that the transition is of the first order. The high-, intermediate-, and low-temperature phases will be called phases I, II, and III, respectively.

Evidence for the phase transition at $T_{s2}$ has been given in the specific heat measurements. They have barely detected a corresponding peak around 120 K, although it is extremely small compared with that found at $T_{s1}$. The peak height is about 0.4 % of the total heat capacity. We could approximately evaluate the entropy change associated with the transition to be 0.04 J/K mol Re, which is only 1 % of that for the phase transition at $T_{s1}$, $\Delta S$ = 3.5 J/K mol Re. This indicates that the reduction of symmetry is much smaller for the transition at $T_{s2}$ than $T_{s1}$.

On the other hand, as shown in Fig. 3, no corresponding anomaly is seen at $T_{s2}$ in magnetic susceptibility measurements. This is in contrast to the large decrease observed at $T_{s1}$, which is not due to magnetic order but due to the reduction of density-of-state (DOS).[6] Thus, the second phase transition should involve a relatively small change in the band structure.

We have measured a transverse magnetoresistance $\Delta\rho$ under magnetic field up to 14 T. It revealed significant anisotropy at $T$ = 2 K possibly reflecting the topology of the Fermi surface and the multi-band character of the band structure,[7, 8] the detail of which will be reported elsewhere. Here we present the temperature dependence of $\Delta\rho$. The measurements were performed at a magnetic field of $H$ = 14 T applied along the [001], [110], or [111] directions of the cubic unit cell. As shown in Fig. 4, it is positive and quite large at low temperature, about 30 % at $T$ = 2 K, along the three directions. On heating, however, it decreases steeply and reaches nearly zero at $T$ = 100 K. Surprisingly, it recovers above 120 K to form a peak around 160 K, and then again disappears above 200 K. To be noted is that the $\Delta\rho$ is anisotropic at the intermediate temperature between 120 and 200 K: positive for [001], almost zero for [111], and negative for [110]. These temperature dependences indicate that a significant change of the Fermi surface occur successively at $T_{s1}$ and $T_{s2}$. Note that the change in DOS occurs only at $T_{s1}$, and is negligibly small at $T_{s2}$.

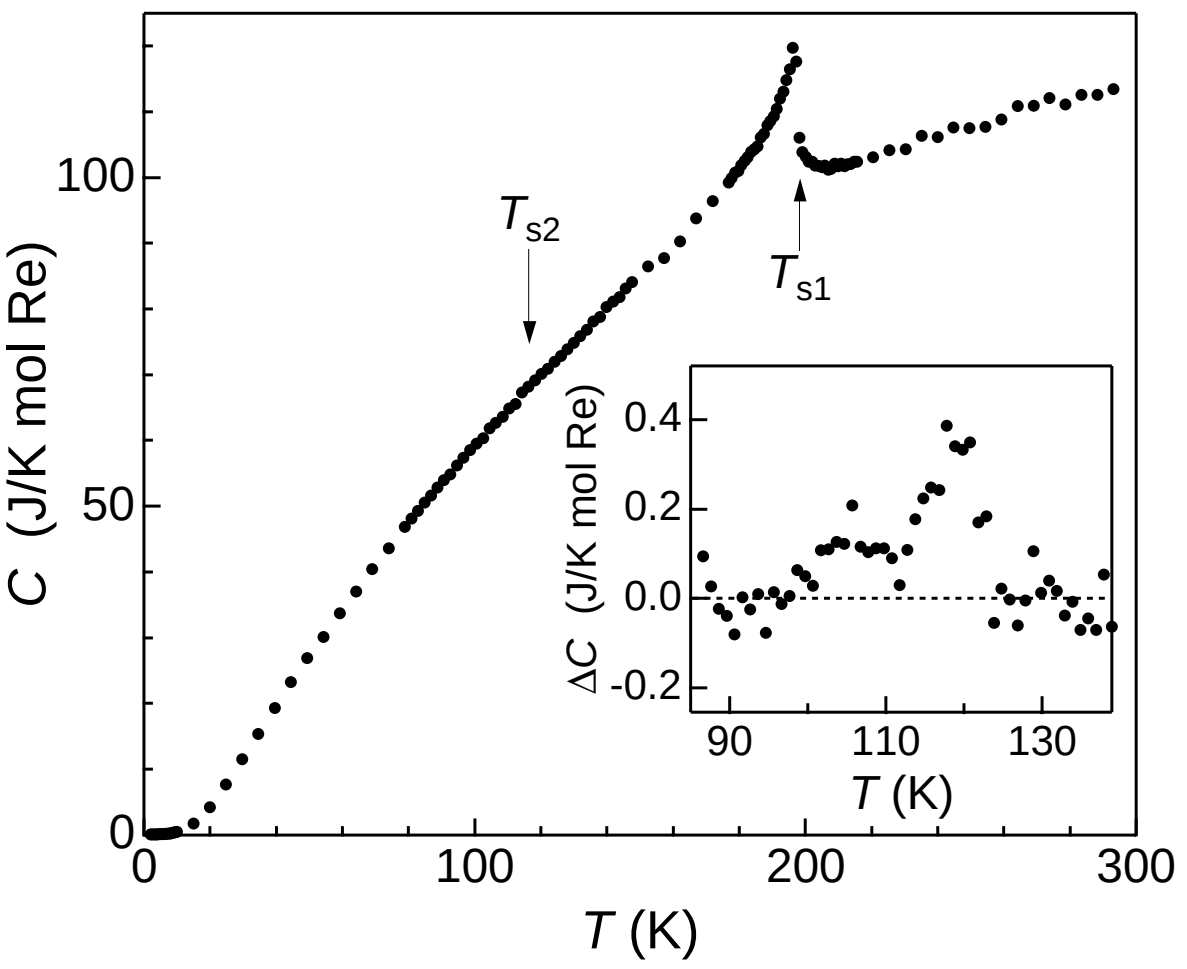

Fig. 2. Specific heat $C$ as a function of temperature. The inset shows the data around 120 K after subtraction of the background.

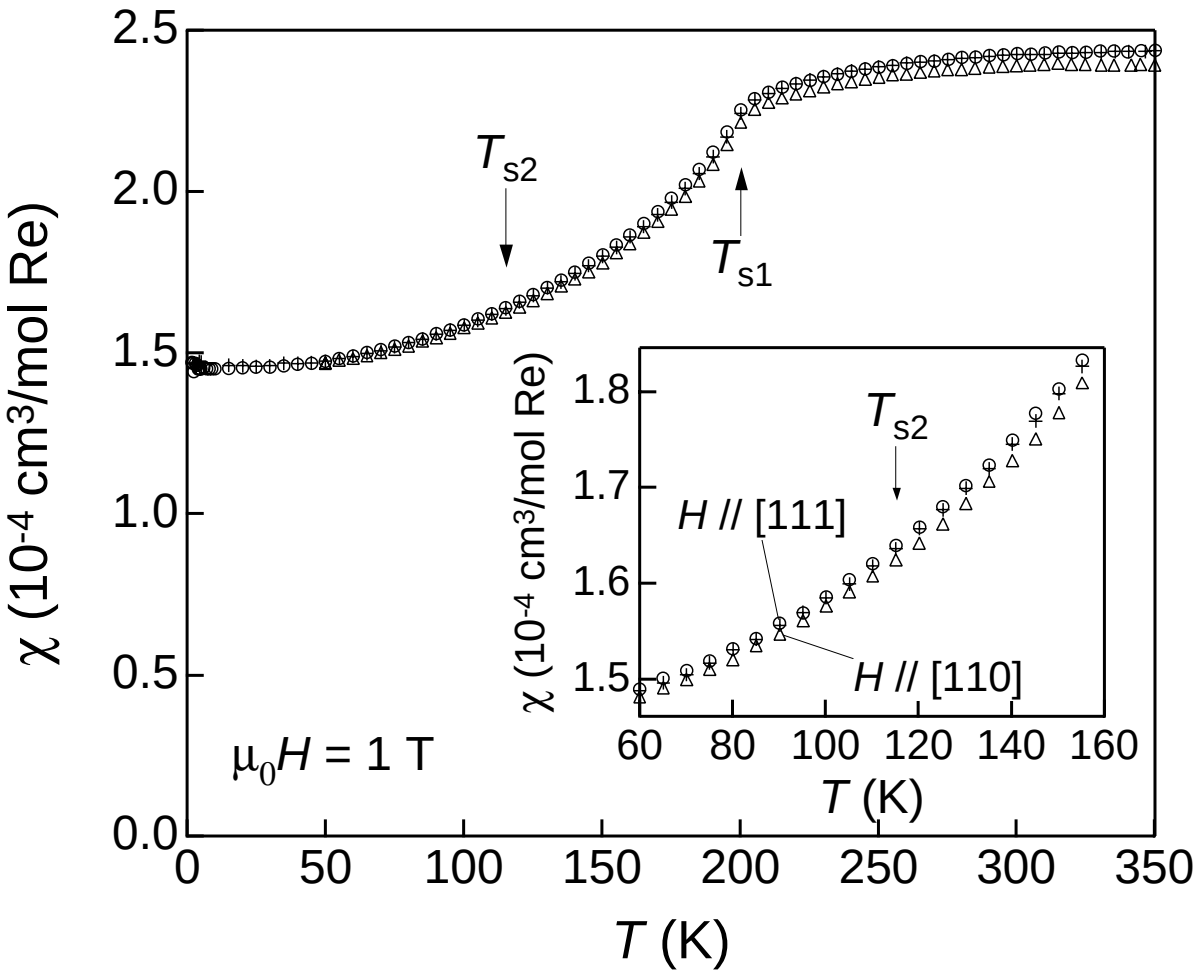

Fig. 3. Magnetic susceptibility $\chi$ measured at a magnetic field of 1 T applied along the [110] and [111] directions of the cubic unit cell. The former was done on heating after zero-field cooling, and the latter on heating (open circle) and then on cooling (cross). No anomalies are detected at $T_{s2}$.

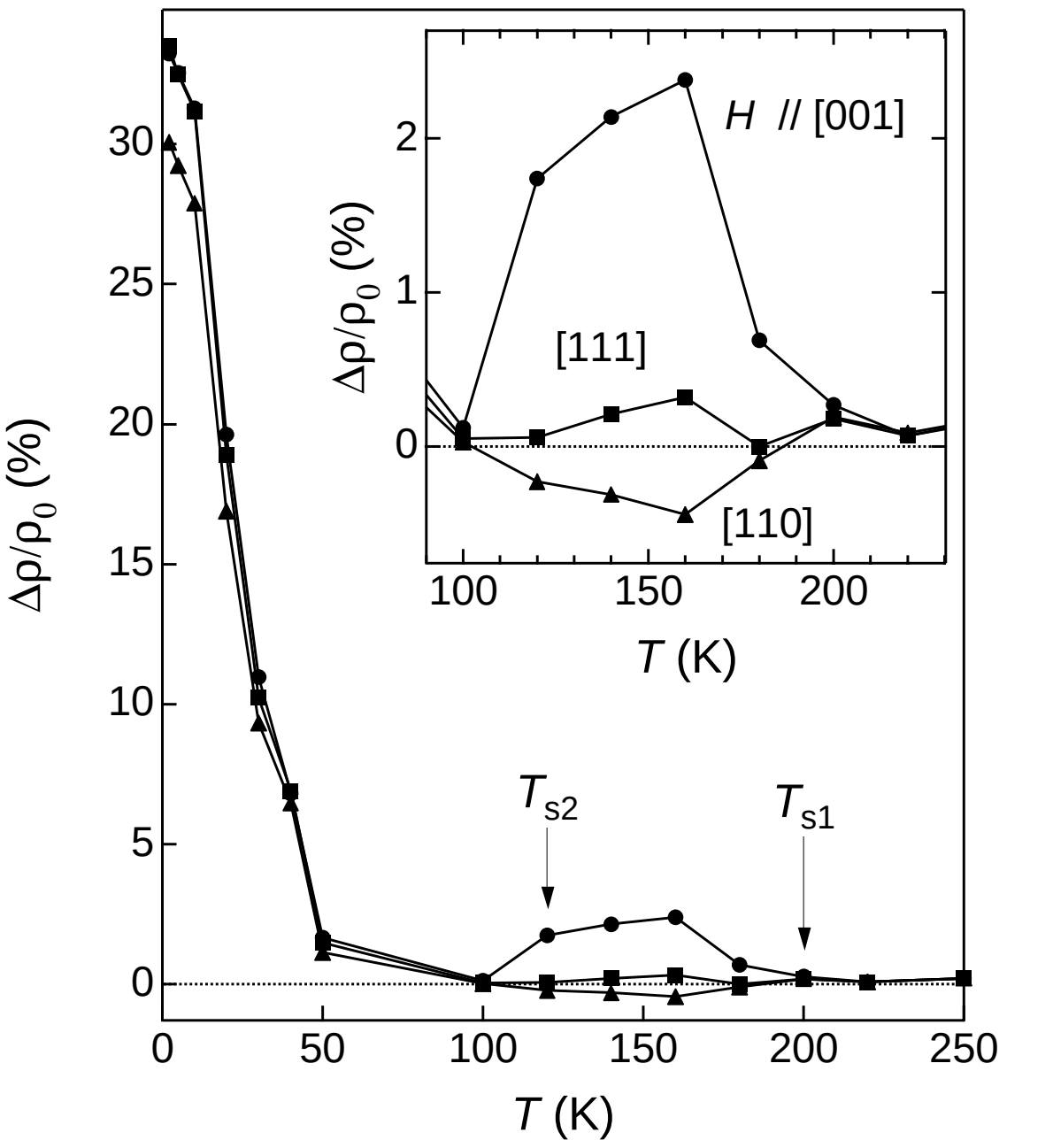

Fig. 4. Magnetoresistance $\Delta\rho$ measured at a magnetic field of 14 T applied along the [001], [110], and [111] directions with electrical current always running perpendicular to the field. Plotted is the temperature dependence of $\Delta\rho$ divided by $\rho_0$, where $\rho_0$ (11.5 μΩcm) is the resistivity at $T$ = 2 K without fields.

Concerning the possibility of structural change at $T_{s2}$, previous XRD experiments on the temperature dependence of the lattice parameter did not detect any anomalies at $T_{s2}$, but a large change at $T_{s1}$.[4] Moreover, single-crystal XRD experiments suggested that a cubic symmetry was preserved down to $T$ = 10 K, and that extra reflections which are compatible with space group *F*43*m* appeared below $T_{s1}$, without any additional ones below $T_{s2}$.[4] In contrast, recent Re NQR experiments indicated a lower symmetry for phases II and III.[6] These facts imply that the structural change at $T_{s2}$ would be very small, if any, compared with that at $T_{s1}$. We have checked again the crystal structure by examining many single crystals down to $T$ = 85 K. Figure 5 gives typical results showing the temperature dependence of intensity for several reflections. Extra reflections (forbidden for space group *Fd*3*m*) like 006 and 00 10 appear below 200 K and gain their intensity gradually, as expected for a second-order phase transition. They show a small anomaly also at 120 K. On the other hand, the intensity of fundamental reflections like 008 and 262 also increases gradually below $T_{s1}$, followed by a notable change at 120 K. Note that the intensity of the 008 reflection exhibits a jump at 120 K, consistent with the first-order transition deduced from the resistivity measurements. These results clearly indicate that structural transitions occur successively in two steps. Further structural analysis is in progress.

In conclusion we have reported experimental evidence for the second phase transition at $T_{s2}$ = 120 K as well as the first one at $T_{S1}$ = 200 K in $Cd_2Re_2O_7$. The second transition is probably of the first order, and is accompanied with a small change in the crystal structure and with a significant change in the electronic structure. As a result, $Cd_2Re_2O_7$ exhibits unique successive structural and electronic phase transitions on cooling: The first one at $T_{s1}$ is characterized by a large reduction in DOS, and the second one at $T_{s2}$ by a change in Fermi surface as detected in magnetoresistance measurements. Interesting physics must be involved there, which is caused by the electronic instability of the itinerant electrons in the pyrochlore lattice.

We have already reported the high-pressure study on $Cd_2Re_2O_7$ where both $T_{s1}$ and $T_{s2}$ are reduced with increasing pressure: $T_{s1}$ disappears around $P$ = 3.5

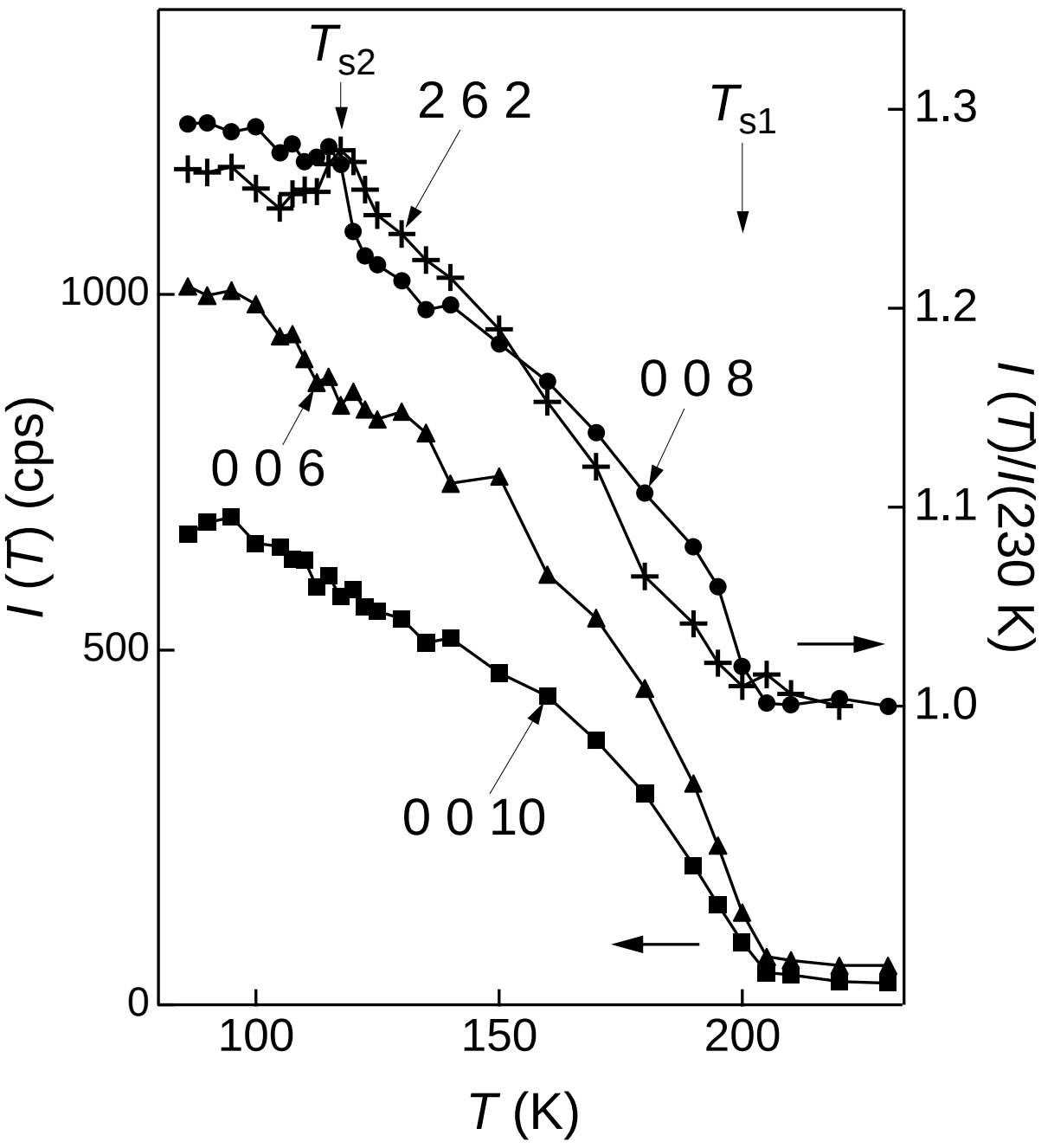

Fig. 5. Temperature dependence of X-ray intensity for several reflections measured using a single crystal. The index is based on the cubic cell. The intensity for the 006 and 00 10 reflections is on the left axis, and that normalized at 230 K for the 008 and 262 is on the right axis.

GPa, while $T_{s2}$ seems to go away at $P$ = 2 GPa in the *P-T* phase diagram. To be noted is that $T_c$ seems to increase with pressure and shows a maximum around $P$ = 2 GPa. It would be very interesting if a fluctuation associated with the second phase transition is relevant to the occurrence of superconductivity in $Cd_2Re_2O_7$.

Acknowledgments

We are grateful to M. Takigawa and H. Harima for helpful discussions. We also thank T. Kamiyama, H. Sawa, K. Tsuda, and S. Sugai for providing us with information on their experimental results. This research was supported by a Grant-in-Aid for Scientific Research on Priority Areas (A) and a Grant-in-Aid for Creative Scientific Research given by The Ministry of Education, Culture, Sports, Science and Technology, Japan.

5